\documentclass{acm_proc_article-sp}

\usepackage{graphicx}
\usepackage{comment}

\begin{document}

% --- Author Metadata here ---
\permission{} % Allows default copyright data (0-89791-88-6/97/05) to be over-ridden - IF NEED BE.
% --- End of Author Metadata ---

\title{GLoP: Enabling Massively Parallel Incident Response Through GPU Log Processing}%

\numberofauthors{5} %  in this sample file, there are a *total*
\author{
%, , Robert C. Atkinson
% 1st. author
\alignauthor
Xavier J. A. Bellekens\\
       \affaddr{Department of Electronic and Electrical Engineering}\\
       \affaddr{University of Strathclyde}\\
       \affaddr{Glasgow, G1 1XW, UK\\xavier.bellekens@strath.ac.uk}
\alignauthor
Christos Tachtatzis\\
       \affaddr{Department of Electronic and Electrical Engineering}\\
       \affaddr{University of Strathclyde}\\
       \affaddr{Glasgow, G1 1XW, UK\\
       christos.tachtatzis@strath.ac.uk}
\alignauthor
Robert C. Atkinson\\
       \affaddr{Department of Electronic and Electrical Engineering}\\
       \affaddr{University of Strathclyde}\\
       \affaddr{Glasgow, G1 1XW, UK\\
       robert.atkinson@strath.ac.uk}
\and  
\alignauthor     
Craig Renfrew\\
       \affaddr{Agilent Technologies UK}\\
       \affaddr{5 Lochside Avenue}\\
       \affaddr{Edinburgh, EH12 9DJ, GB\\
       craig\_renfrew@agilent.com}
\alignauthor
Tony Kirkham\\
       \affaddr{Agilent Technologies UK}\\
       \affaddr{5 Lochside Avenue}\\
       \affaddr{Edinburgh, EH12 9DJ, GB\\
       tony\_kirkham@agilent.com}
}
% There's nothing stopping you putting the seventh, eighth, etc.
% author on the opening page (as the 'third row') but we ask,
% for aesthetic reasons that you place these 'additional authors'
% in the \additional authors block, viz.

% Just remember to make sure that the TOTAL number of authors
% is the number that will appear on the first page PLUS the
% number that will appear in the \additionalauthors section.

\maketitle
\begin{abstract}
Large industrial systems that combine services and applications, have become targets for cyber criminals and are challenging from the security, monitoring and auditing perspectives. Security log analysis is a key step for uncovering anomalies, detecting intrusion, and enabling incident response. The constant increase of link speeds, threats and users, produce large volumes of log data and become increasingly difficult to analyse on a Central Processing Unit (CPU). This paper presents a massively parallel \textbf{G}raphics Processing Unit (GPU) \textbf{Lo}g \textbf{P}rocessing (GLoP) library and can also be used for Deep Packet Inspection (DPI), using a prefix matching technique, harvesting the full power of off-the-shelf technologies. GLoP implements two different algorithm using different GPU memory and is compared against CPU counterpart implementations. The library can be used for processing nodes with single or multiple GPUs as well as GPU cloud farms. The results show throughput of 20~Gbps and demonstrate that modern GPUs can be utilised to increase the operational speed of large scale log processing scenarios, saving precious time before and after an intrusion has occurred.
\end{abstract}

\category{D.4.6}{Security and protection }[Information flow controls]

 % NOT required for Proceedings
\\

\section{Introduction}
Incident Response and threat detection play a key role in industrial system security. With the increased dependence of users on information gathering, e-commerce, social networking, and the Internet of Things (IoT), the amount of data to analyse before and after an intrusion has grown exponentially. This large data volume, in combination with the expanding number of malicious exploits from attackers, make it challenging for system administrators and incident response teams to analyse the logs. Existing frameworks and tools utilise modern search algorithms, however, most of these run sequentially on CPUs.

Research in malware detection~\cite{5437720},~\cite{5174096}, Intrusion Detection Systems (IDS)~\cite{4041182},~\cite{FPGA1}, and pattern matching~\cite{1364659},~\cite{6299287} demonstrate significant speed increases utilising parallelised hardware architectures such as Field Programmable Gate Arrays (FPGA) and Graphical Processing Units (GPU). These approaches effectively scale processing vertically and maximise speed from a single device. Cloud based research~\cite{shu},~\cite{6045076}, attempts to parallelise log-processing horizontally, by distributing workload to multiple nodes with services such as the Amazon EC2, GPU G2 instances. These two approaches, are by no means competing but rather complementary. Combining GPU processing in the cloud has the potential of massive speed increases.

This paper addresses the analysis of large scale log processing in a fast and cost-effective fashion using a single off-the-shelf GPU. The performance increase is not limited to a single GPU and can be utilised to enhance both multi-GPU nodes as well as GPU Cloud farms.

\begin{figure}[!tb]
\centering
\includegraphics[width=3.5in, height=1.8in]{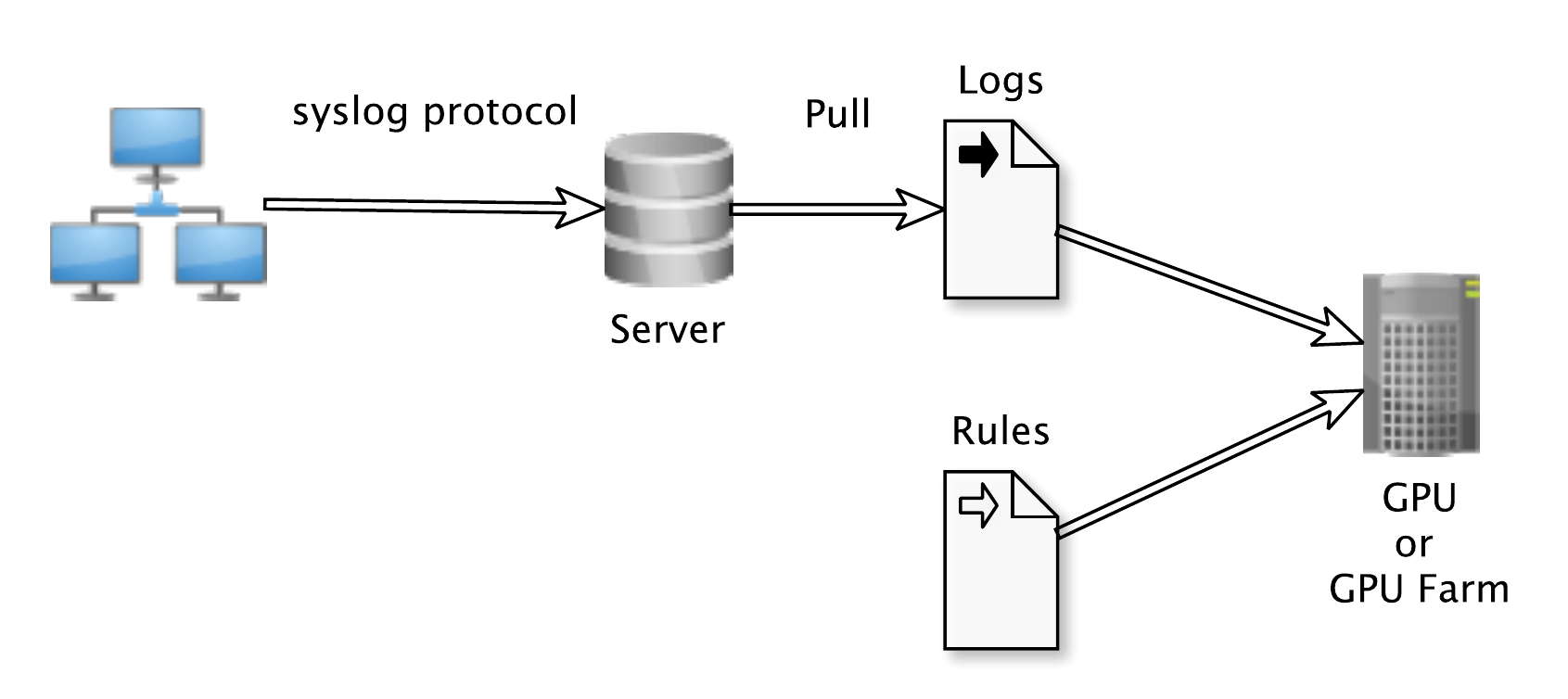}
\caption{Typical use case scenario}
\label{fig:usecase}
\end{figure}

The typical use case scenario of the library is illustrated in Figure~\ref{fig:usecase}. Network equipment record events using the syslog protocol~\cite{rfc5424} on a server. The incident response team would periodically push the syslog traces for the last period, to a GPU processing server along with rules describing malicious activity. The GPU server will search through the logs and identify patterns that match any of the malicious rules generating alerts containing the number of suspected malicious activities and their location in the logs.

The contributions of the paper are threefold. It proposes and implements a generic log processing library, called GLoP (\textbf{G}PU \textbf{Lo}g \textbf{P}rocessing). The library utilises the single-pattern matching algorithm Knut-Morris-Pratt (KMP) and the multi-pattern algorithm of Failureless Tries. Both algorithms are implemented for CPU and GPU and the performance of the implementations is quantified, for various number of search patterns. The library truncates the patterns to 8 characters reducing thread divergance and memory requirments on the GPU; a technique known as prefix or partial matching. Finally, GLoP contains two GPU implementations of Parallel Failureless Aho-Corasick algorithm that use global or texture memory. It is shown, that the implementation using the texture memory achieves double throughput compared to the implementation that uses global memory.

The remainder of this paper is organised as follows: Section~\ref{sec:related_work} presents the related work, Section~\ref{sec:cuda_programming_model} introduces the GPU architecture and programming model. Sections~\ref{sec:aho} and~\ref{sec:PFAC} describe the Aho-Corasick algorithm and optimisation for its implementation in GLoP. The experimental environment and methodology are presented in Section~\ref{sec:experimental_environment}. Section~\ref{sec:evaluation} describes and discusses the results obtained from multiple algorithm variants within GLoP. Finally, Section~\ref{sec:conclusion} presents the conclusions and future work.

\section{Related Work}
\label{sec:related_work}
Large scale log processing research has extensively studied the use of large scale data mining and big data scenarios using distributed frameworks analysis. \emph{Shu et al.} have proposed a lightweight framework based on the Amazon Cloud Environment (EC2 and S3), using multiple nodes to speed up the log analysis processing, and harvesting the results using a map reduce implementation~\cite{shu}. \emph{Yang et al.} demonstrated that by using Hadoop MapReduce, it was possible to decrease the processing time of log files by 89\% for intrusion detection purposes~\cite{6045076}. 
\emph{Marty et al.} proposed a theoretical logging framework dedicated to cloud infrastructures and software as a service (SaaS) running on a third party public cloud service~\cite{Marty:2011:CAL:1982185.1982226}.

\emph{Cheng et al.} described a fast filter virus detection engine running on GPUs based on eigenvalues with good performances~\cite{5437720}. Our previous work~\cite{bellekens} has shown that a massively parallel pattern matching algorithm based on the Knuth-Morris-Pratt algorithm~\cite{doi:10.1137/0206024} can achieve a 29 fold increase in processing speed over CPU counterparts. From the output of these works, it is clear that the processing capabilities of off-the-shelf hardware have a great potential not only to increase the speed on a single stand-alone processing server but also on GPU cloud deployments \cite{6227786}.

String searching algorithms can be classified to single-pattern and multi-pattern matching. Single pattern matching algorithms search the complete string for a single pattern sequentially. The naive approach to search for a single pattern is to iteratively walk through the text string and every time there is a mismatch or a complete match, the algorithm rewinds back. Optimised algorithms such as the Knuth-Morris-Pratt (KMP) and Boyer-Moore (BM) avoid rewinding by introducing failure and backtracking tables respectively~\cite{Boyer:1977:FSS:359842.359859}.

On the other hand multi-pattern matching algorithms search simultaneously for multiple patterns in the text string. The most common multi-pattern algorithm is the Aho-Corasick (AC)~\cite{Aho:1975:ESM:360825.360855}, \cite{Tan:2005:HTS:1069807.1069981} which has been implemented in a variety of hardware architectures such as FPGAs~\cite{FPGA1}\cite{4285750} and GPUs~\cite{6165263}.

For cases where the cross-correlation between pattern prefixes is low, a two staged searching approach can be utilised to improve parsing speeds. In these scenarios, the patterns are truncated to create a set of prefixes. The first searching stage filters the text locations where the prefixes match. Subsequently the identified locations are passed to a secondary process where the patterns beyond the prefix are searched. This completes the searching for the full pattern. Such techniques are known as prefix or partial pattern matching~\cite{5928778, 5440330}. In this work, only the first searching stage is considered as the prefix cross-correlation is low and patterns share the same prefix for 0.0001\% of the time~\cite{Gravity}. Therefore the additional overhead to complete the full pattern search is negligible.

\section{CUDA Programming Model}
\label{sec:cuda_programming_model}
The Compute Unified Device Architecture (CUDA) framework offers the possibility to researchers to use GPUs for General Purpose Graphics Processing. The framework allows researchers to access hardware features using an extension of the C99/C++ language from the host. 

The present implementation of the library is implemented for the Tesla graphics card. The device consists of different multiprocessors each one of them containing streaming processors (SP). Each SP executing thousands of threads. Threads running on the GPU are organised in thread blocks. Within each block, the threads are organised in warps, each warp contains the same number of threads (32 for the Tesla K20 GPU). Each wrap executes in a Single Instruction Multiple Thread (SIMT) fashion and the multiprocessor periodically switches between warps maximising the resources, and hiding latencies~\cite{kirk2012programming}. 

The GPU contains different types of memory, \textit{Global, Texture, Constant, Shared} which need to be managed explicitly at compilation time, by the programmer. The global memory is the largest memory on the device, and requires the more clock cycles to be accessed. The texture and constant memories are cached, and require less clock cycles to be accessed. Finally the shared memory is shared between threads within a block and requires even fewer clock cycles but is a very scarce resource~\cite{Nvidia}.

\section{Aho-Corasick}
\label{sec:aho}
The AC algorithm is a multi-pattern matching algorithm and is able to identify multiple patterns in the text string on a single pass. The AC algorithm consists of two phases: the first phase constructs a state machine from a set of patterns, while the second phase searches for the patterns in the text.

\begin{figure}[!b]
\centering
\includegraphics[width=3.62in]{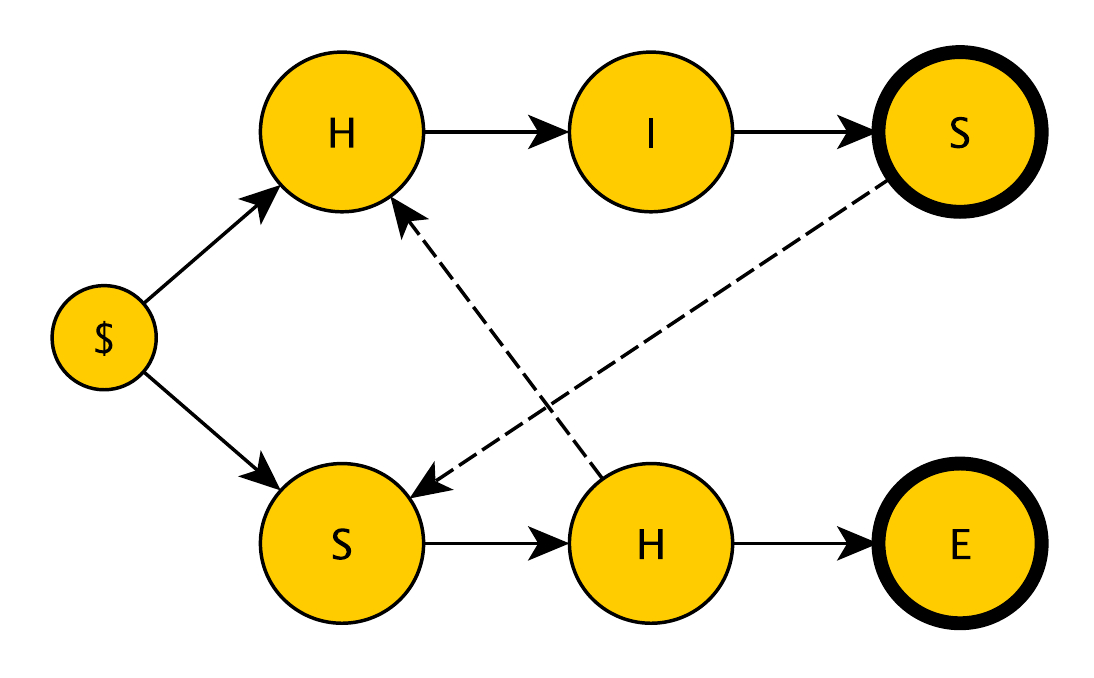}
\caption{Aho-Corasick}
\label{fig_AC}
\end{figure}

A typical state machine diagram constructed during the first phase of the algorithm is shown in Figure~\ref{fig_AC}. The state machine shown, consists of two patterns $P$~=~$\{HIS, SHE\}$. The straight arrows represent the ``regular'' transitions which occur when the text is matching, while the dashed arrows represent the ``failure'' transitions, when a mismatch occurs but the prefix of another pattern is matched. This allows single threaded implementation of the AC algorithm to avoid rewinding and back-track matches of multiple patterns on a single pass.

For example, for the text $T=``SHIS''$ and the state machine of Figure~\ref{fig_AC}, the algorithm will start with two consecutive regular transitions from the initial state \$ to $S$ and from $S$ to $H$ when comparing the first two characters of the text $T$. On the third comparison, the character $I$ from the text string $T$ does not result to a regular transition and consequently the failure transition is activated followed by a regular transition from $H$ to $I$. Finally, for the last comparison results to a regular transition from $I$ to $S$.

\begin{figure}[!t]
\centering
\includegraphics[width=3.5in]{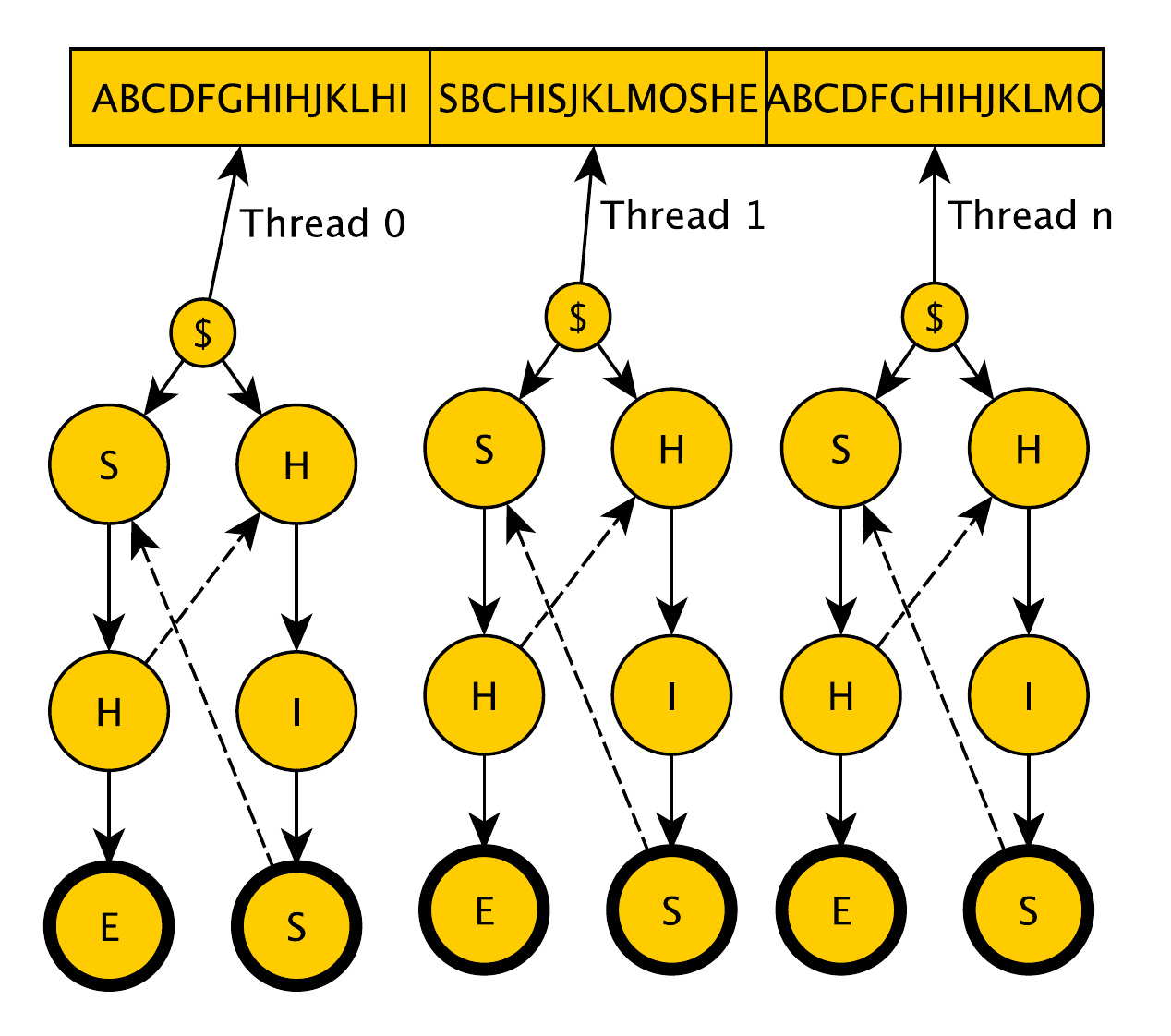}
\caption{Aho-Corasick Boundary Problem}
\label{fig_boundary}
\end{figure}

The most common modification to adopt the AC algorithm in a multi-threaded environment is to split the text $T$ in multiple chunks and assign one chunk per thread. However, this creates problems when the text contains patterns across two chunks. For example, when the state machine of Figure~\ref{fig_AC} is considered, for the text string shown in Figure~\ref{fig_boundary}, it is clear that the character sequence $``HIS\textrm'\textrm'$ exists on the boundary between the chunks assigned to \emph{Thread 0} and \emph{Thread 1} and neither of these threads will match it. That can be avoided if the text string is split in chunks that overlap each other by at least the length of the longest pattern minus 1. The implication of the minimum chunk overlap is that it limits the number of chunks especially for scenarios where the maximum pattern length is long. Consequently the number of threads used to process the input text are also limited and this could become a bottleneck considering the massive number of threads that can be launched in the GPU (tens of thousands)~\cite{5290944}.

Another drawback of the AC algorithm in a multi-threaded environment is that the failure links create thread divergence. Thread divergence occurs when threads of the same wrap branch to different sections of the code. As described in Section~\ref{sec:cuda_programming_model}, the GPU executes the wraps in a Single Instruction Multiple Thread manner and the GPU scheduler will allocate an execution slice to every branch. Threads on the same branch execute concurrently, while all the threads on a different branch are waiting for an execution slice. The scheduling within the wrap occurs in a Round Robin fashion. Thread divergence does not only consume more execution slices, but also increases memory access time as each branch makes memory transfers at different execution slices making access non-coalesced.

% THE FOLLOWING TWO PARAGRAPHS WHERE THE ORIGINAL DESCRIPTION
%In the field of pattern matching, the Aho-Corasick (AC) algorithm has been widely used~\cite{Aho:1975:ESM:360825.360855}, \cite{Tan:2005:HTS:1069807.1069981} due to its advantage of matching a set of patterns in a single pass. The AC algorithm is composed of two specific phases. The first phase consists of merging a set of patterns into a single state machine, while the second phase consists of using a single CPU thread to match this set of patterns against a text. Figure~\ref{fig_AC} is composed of two patterns $P$~=~$\{HIS, SHE\}$. The straight arrows are representing the valid transitions, while the dashed arrows are representing the failure links. 

%Failure links allow a single thread to back-track different patterns in a single pass. (E.g. Matching the text $HISHE$ is possible using a single thread and one single pass). On the GPU the AC failure links imply more divergence during the pattern matching, within each thread following a different path. To solve the problems involved in the AC algorithm on a GPU \emph{Lin et al.} proposed a Failureless Trie~\cite{6338923}. 

\section{Parallel Failureless Trie and \\ Prefix Matching}
\label{sec:PFAC}
\begin{figure}[!b]
\centering
\includegraphics[width=3in]{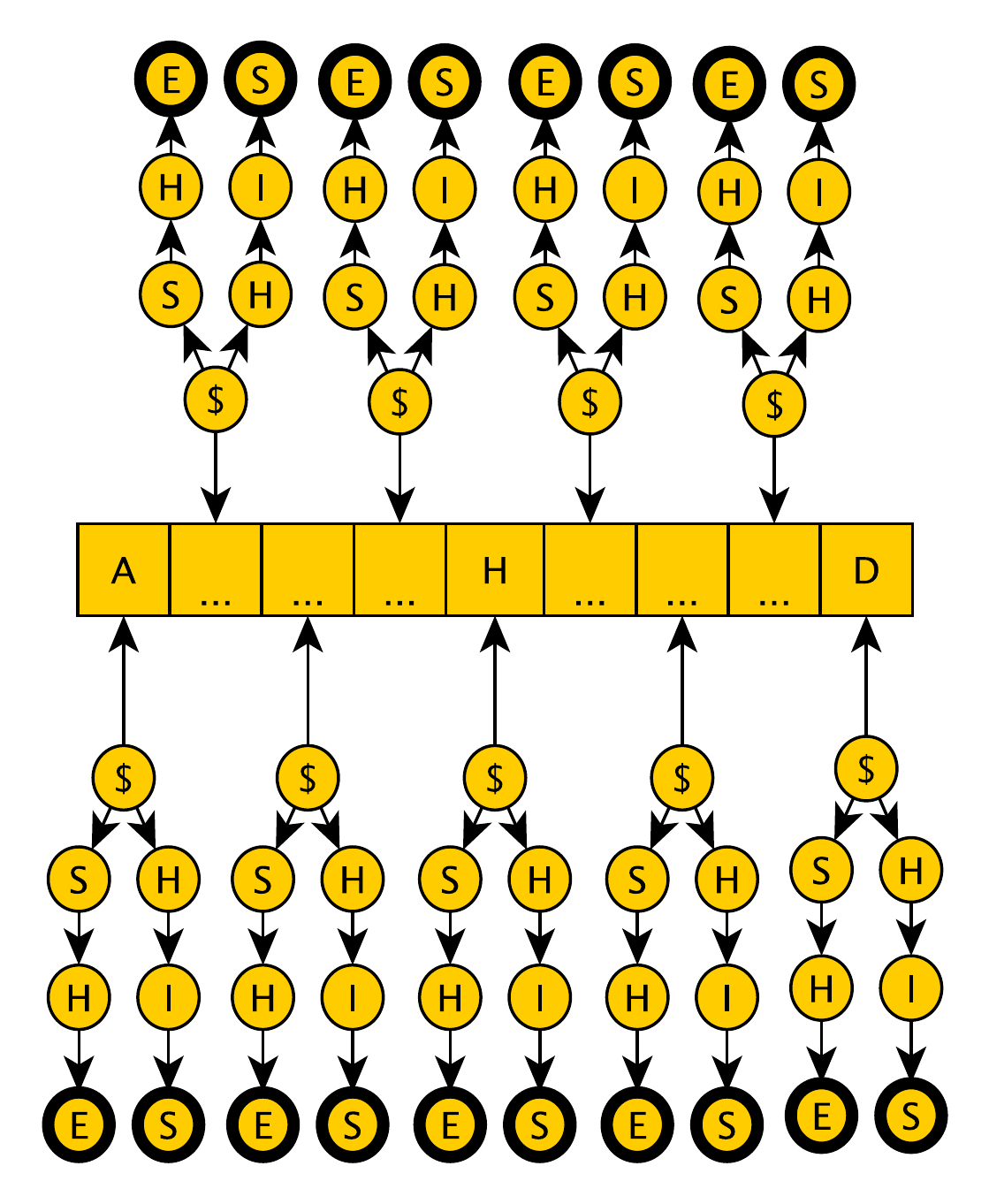}
\caption{One Thread per Byte}
\label{fig_thread}
\end{figure}

An alternative method to parallelise the AC algorithm that avoids the problem of patterns splitting over boundaries and reduces thread divergence is the Parallel Failureless Aho-Corasick (PFAC)~\cite{6338923}. With this method, ``failure'' transitions are removed from the state machine and one thread is instantiated for every character in the text string as shown in Figure~\ref{fig_thread}. In the case of a match, the current thread will overlap the succeeding thread, allowing the state machine to change state. In the case of a mismatch, the current thread will terminate, releasing the resources and reducing divergence.

\begin{figure*}[!t]
\centering
\includegraphics[width=6in]{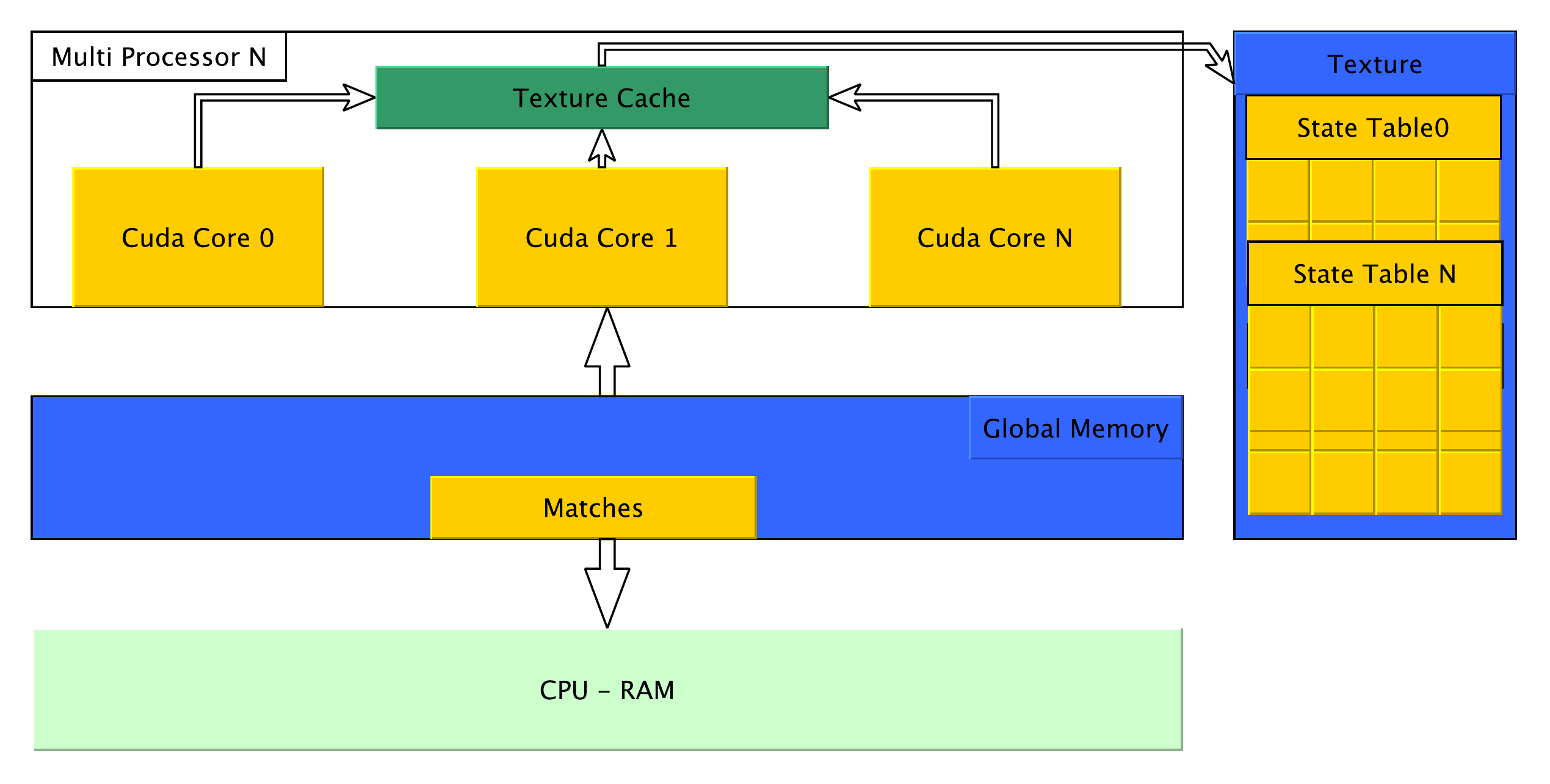}
\caption{Texture Memory Architecture}
\label{fig_arch}
\end{figure*}

The failureless trie mechanism exploits effectively the large number of threads that are available in massively parallelised off-the-shelve GPU devices, however further optimisation can be achieved.

Observe that the first memory transfer of text string to the threads occurs in a coalesced manner. Subsequently, each thread will transition to separate branches of the trie, leading to non-coalesced memory access and thread divergence. As the algorithm progresses deeper in the trie more threads will terminate due to mismatches leaving only a small number of threads continuing to match while the bulk of the threads terminate. This reduces the throughput significantly and can be avoided if the trie is truncated and all the threads terminate beyond a certain trie depth. The approach of truncating the trie is known as prefix or partial pattern matching~\cite{938073, Gravity} and assumes that a secondary process takes over to complete the search; for example a CPU thread is lunched to complete the search.

In practise this approach is effective as the prefix length of a malicious activity pattern has a low probability being a false positive (i.e. the prefix exist in the text followed by a suffix that does not match that pattern). \emph{Vasiliadis et al.}~\cite{Gravity} has shown that a prefix length of 8 is adequate to achieve a false positive rate below 0.0001\%. In this work this has been verified independently.

The prefix matching does not only reduce divergence, but also reduces memory footprint due to the reduced number of states in the trie. Memory reduction in itself may not be of significant benefit, but when texture memory is used the caching hit rate is increases leading to significantly reduced memory access times.

\section{Experimental Environment}
\label{sec:experimental_environment}
GLoP was implemented on a Nvidia Tesla K20m graphic card. The card consists of a single Printed Circuit Board (PCB), composed  of 192 CUDA Cores distributed over 19 multiprocessors. The card is also composed 5~GB of global memory,and permits up to 26,624 active threads. 

The base systems, is supplied with two Intel Xeon E5-2620 six core CPUs running at 2.0 GHz. The rig has a total of 64~GB ram. The server is running Ubuntu Server 11.10 (kernel version 3.0.0-12-server). The algorithm has been compiled with the optimisation brought by the latest version of CUDA 6.0, using C99.

\begin{figure}[!t]
\centering
\includegraphics[width=3.5in]{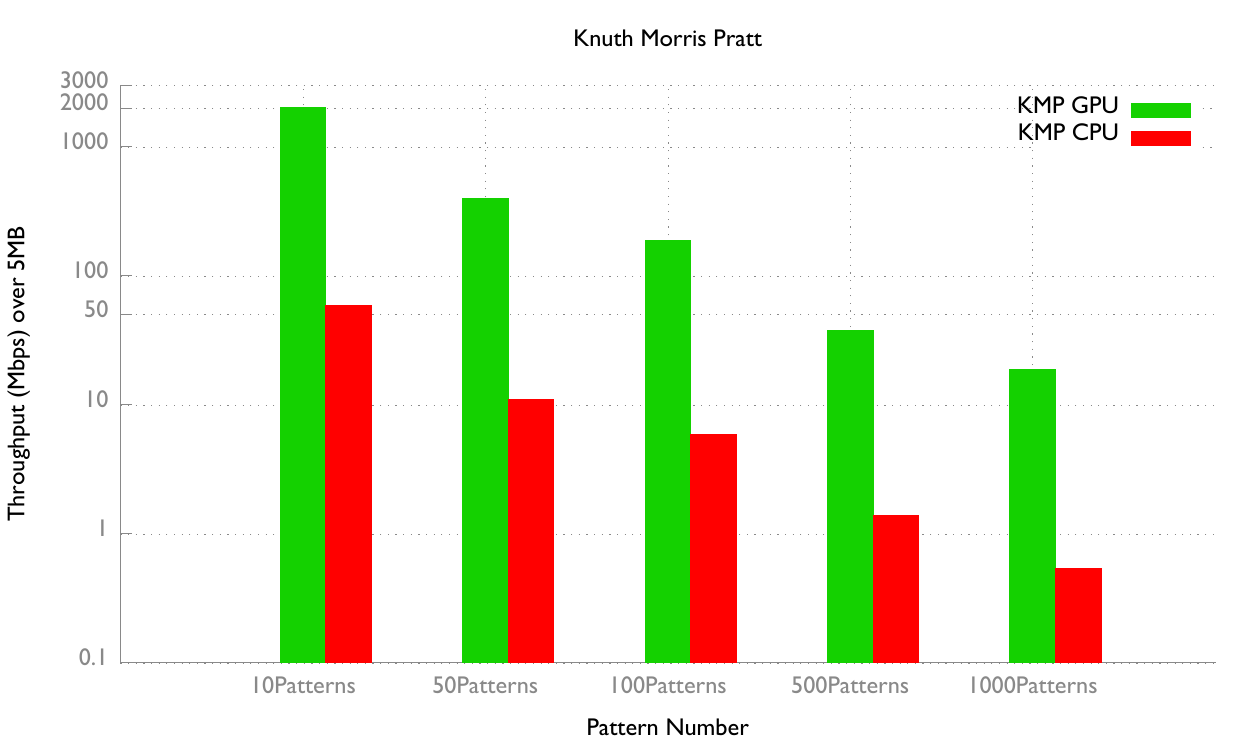}
\caption{Throughput of the KMP Algorithm over 5MB chunk of log files}
\label{fig_KMP}
\end{figure}

The GLoP library is executed using unified memory, allowing the users to access variables stored in global memory from the CPU and the GPU at any given time, and to facilitate the execution of streamed kernels. Each stream sends logs to GPU to be analysed, and matched against the failureless trie. During the launch of a kernel, the CPU is free to stream the next batch of logs to the GPU.

Two implementations of the algorithm are evaluated using global and texture memory. For the texture memory versions, the trie must be represented in memory using \texttt{int} types which are subsequently matched to ASCII values (e.g. ``A''=65). The texture memory allows transactions to be cached, requiring fewer clock cycles to complete transfers compared to global memory accesses. The caching procedure is illustrated in Figure~\ref{fig_arch} where CUDA Cores (within a Streaming Multiprocessor) have access to texture memory via the cache, potentially speeding up significantly memory access.

During the evaluation, the algorithm is run 100 times and the results presented are the average of the total runs. This allows to mitigate the sources of jitter; such as background processes running competing for resources available. The patterns searched are also generated randomly against each log files. 

The log files used during the trials have been automatically generated, using the Mersenne Twister (MT) uniform pseudo-random number generator~\cite{Matsumoto:1998:MTE:272991.272995}. Each synthetic log has an exact file size of 100~MB, and its uniqueness is ensured by computing the SHA256 hash.

In all comparison the throughput is calculated as follows:
\begin{itemize}
\item Let $N$ be the size of the log data sent to the GPU.  
\item Let $Time_{gpu}$ be the time elapsed during with the algorithm is running.
\end{itemize}
\begin{equation}
\cfrac{8 * N}{Time_{gpu}} = Throughput
\label{eq:throughput}
\end{equation}

\section{Evaluation}
\label{sec:evaluation}

\begin{figure}[!t]
\centering
\includegraphics[width=3.5in]{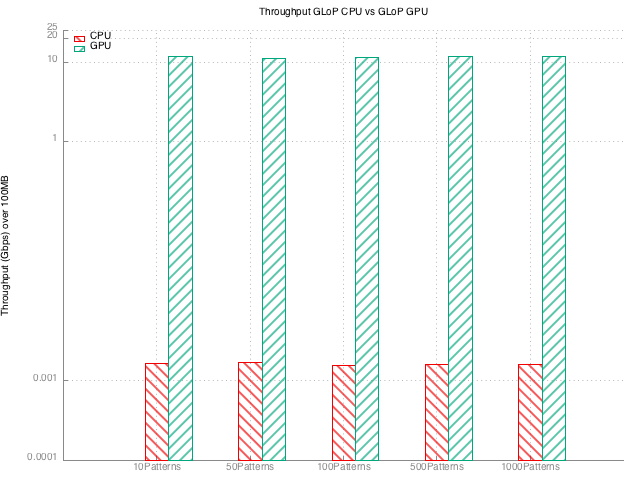}
\caption{Throughput of the GLoP CPU-GPU Library over 100MB log files}
\label{fig_FLTCPU}
\end{figure}

\begin{figure}[!b]
\centering
\includegraphics[width=3.7in]{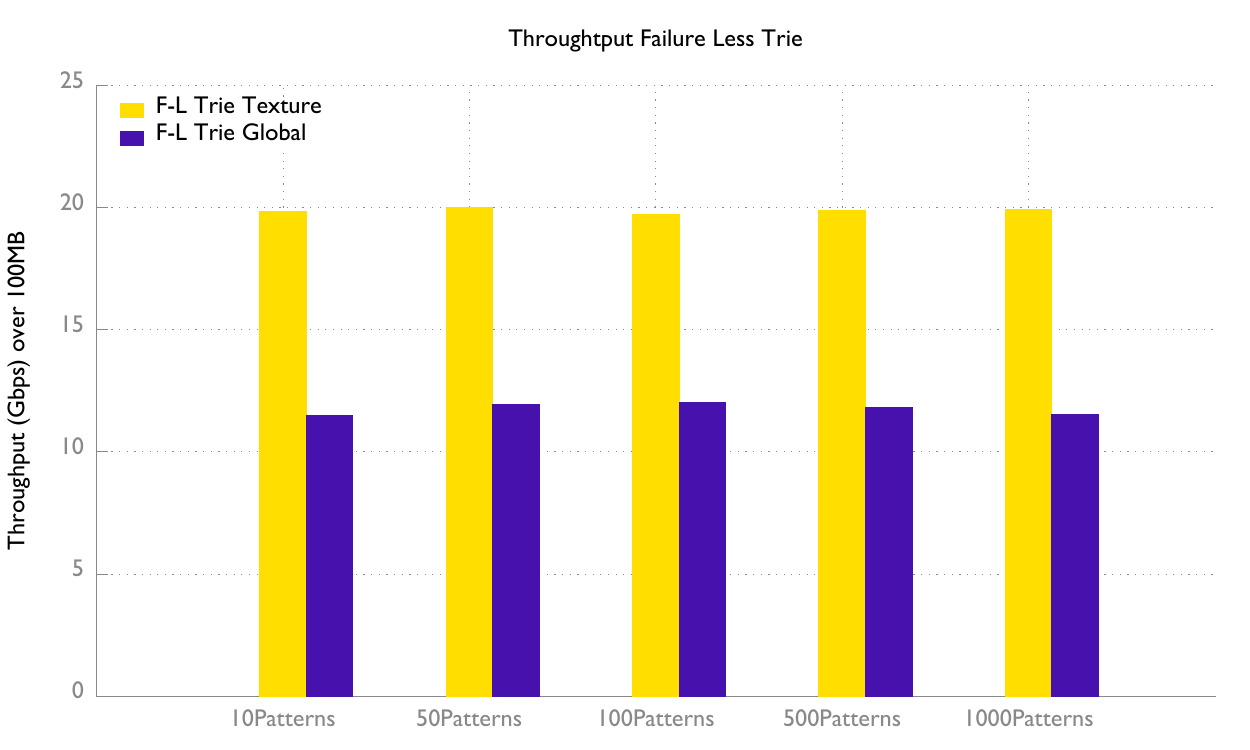}
\caption{Throughput of the GLoP Library over 100MB log files}
\label{fig_FLT}
\end{figure}

\begin{comment}

Two failureless trie algorithms are evaluated, the first one is an implementation of the algorithm only using the global memory, while the second implementation fully harvest the power of the texture memory. In both implementation each thread is assigned to single byte of the text string, starting the matching against the automaton stored either in global memory or texture memory demonstrating the respective performances. 
\end{comment}

GLoP implements the multi-pattern failureless trie algorithm using global and texture memory as well as the single pattern algorithm Knuth-Morris-Pratt for comparison purposes. The KMP algorithm performance on CPU and GPU for varying number of patterns is shown in Figure~\ref{fig_KMP}. As the number of pattern increase, the throughput decreases linearly as the algorithm performs a single pattern search at time. The single threaded CPU version of the KMP algorithm achieves throughput of 50~Mbps when matching a log file of 5~MB against 10 patterns. On the CPU, the throughput is 10 times less when matching the same log file against 100 patterns. Despite the fact that the GPU version of the algorithm achieves throughput nearly 2,000~Mbps when 10 patterns at searched while the throughput decrease linearly to 15~Mbps for 1,000 patterns.

The KMP algorithm does not have sufficient throughput to cope with large scale log analysis, even with the boosted performance on the GPU. Its main shortcoming is that the single pattern searching does not scale well for multiple patterns. GLoP shows that the multi-pattern algorithms such as the failureless trie are able to maintain a constant throughput independent of the number of patterns. This is demonstrated on Figure~\ref{fig_FLTCPU} where the performance of the Parallel Failureless Aho-Corasick algorithm is shown to sustain throughput for both the CPU and GPU. The performance of the failureless trie algorithm on the GPU when global memory is used is 11~Gbps for the 10 patterns and 100~MB file, while the corresponding CPU version achieves throughput of 1.7~Mbps.

Implementation of the failureless trie can be further improved when texture memory is used. The operation of the algorithm in this case is identical with the only different being that the state transition table being stored in texture memory. The performance comparison between the two algorithm using the global and texture memory respectively is shown in Figure~\ref{fig_FLT}. The texture implementation is shown to achieve double the throughput and this can be attributed to the caching capabilities which significantly reduce the cycles required to retrieve data from memory. The global memory implementation culminates at 11~Gbps for a 100~MB log file whereas the texture implementation reaches an overall throughput of 20~Gbps.

\section{Conclusion and Future Work}
\label{sec:conclusion}
This paper presented GLoP, a massively parallel incident response engine offloading the large scale log processing, and pattern matching to the GPU allowing the CPU to concentrate on other tasks.

In this work GLoP has been evaluated against the single pattern matching algorithm Knuth-Morris-Pratt, and has demonstrated a throughput of 11~Gbps with the failure-less global memory implementation and an overall throughput of 20~Gbps for the failure-less texture memory implementation. Various synthetic log files have been used proving its efficacy, and in particular its ability to considerably speed up incident response processes, while remaining cost-effective. 
This work also highlighted the high performances demonstrated by the engine, as a single node, but also its possible and complementary use to cloud based log processing frameworks, such as Hadoop, to increase the processing power. 

In future work, it is planned to reduce the memory footprint of the failure-less trie when using larger state machines, and improve the library by using Message Passing Interface (MPI) technique to allow the library to run on an HPC as well as investigating real-time operations.

\section*{Acknowledgment}
The authors would like to thank Agilent Technologies for their insightful comments and feedback as well as their support. 


\bibliographystyle{ieeetr}
% The original style is abbrv which does not make refernces to appear in order.
% ACM has its own style: \bibliographystyle{acm}
% \bibliographystyle{abbrv}
\bibliography{bibliography}  % sigproc.bib is the name of the Bibliography in this case
% ACM needs 'a single self-contained file'!
%\balancecolumns
\end{document}